\documentclass[review]{elsarticle}

\usepackage{lineno}
\usepackage[colorlinks,
            linkcolor=red,
            anchorcolor=red,
            citecolor=red
            ]{hyperref}
\modulolinenumbers[5]

\journal{Communications in Nonlinear Science and Numerical Simulation}

\usepackage{longtable}
\usepackage{graphicx}
\usepackage{subfigure}
\usepackage{url}
\usepackage{booktabs}
\usepackage{tabularx}
\usepackage{amsthm}
\usepackage{multirow}
\usepackage{enumerate}

\usepackage[]{changes}
  \definechangesauthor[name={Tao Wen}, color=orange]{add}
\usepackage{soul}
  \soulregister\cite7 
  \soulregister\ref7 

\newtheorem{definition}{Definition}[section]

\begin{document}

\begin{frontmatter}

\title{Vital Spreaders Identification in Complex Networks with Multi-Local Dimension}

\author[address1]{Tao Wen}
\author[address2]{Danilo Pelusi}
\author[address1]{Yong Deng \corref{label1}}

\address[address1]{Institute of Fundamental and Frontier Science, University of Electronic Science and Technology of China, Chengdu, 610054, China}
\address[address2]{Faculty of Communications Sciences, University of Teramo, Italy}
\cortext[label1]{Corresponding author at: Institute of Fundamental and Frontier Science, University of Electronic Science and Technology of China, Chengdu, 610054, China. E-mail: dengentropy@uestc.edu.cn, prof.deng@hotmail.com.(Yong Deng)}

\begin{abstract}
  The important nodes identification has been an interesting problem in this issue. Several centrality measures have been proposed to solve this problem, but most of previous methods have their own limitations. To address this problem more effectively, multi-local dimension (MLD) which is based on the fractal property is proposed to identify the vital spreaders in this paper. This proposed method considers the information contained in the box and $q$ plays a weighting coefficient for this partition information. MLD would have different expressions with different value of $q$, and it would degenerate to local information dimension and variant of local dimension when $q = 1$ when $q = 0$ respectively, both of which have been effective identification method for influential nodes. Thus, MLD would be a more general method which can degenerate to some exiting centrality measures. In addition, different with classical methods, the node with low MLD would be more important in the network. Some comparison methods and real-world complex networks are applied in this paper to show the effectiveness and reasonableness of this proposed method. The experiment results show the superiority of this proposed method.
\end{abstract}

\begin{keyword}
    Complex network, Vital spreaders identification, Multi-local dimension
\end{keyword}

\end{frontmatter}


\section{Introduction} \label{Sec_Introduction}

The complex network has become a hot topic in recent research, because it is inextricably correlated with various research issues. For example, the Cyber-Physical Systems (CPS) can be transformed into complex network to study the system operation \cite{Guan2018Internet}, optimization \cite{Xu2012Performance,Guan2010Microgrid}, and reliability \cite{Wei2018Measuring} issues. The traffic network can also use complex networks to study traffic congestion \cite{XU2017AMM}, path planning \cite{Wu2019TDPP}, intelligent transportation \cite{yang2019network,YANG2019121259method}, et al. Therefore, the study of the basic property of complex networks has become more important \cite{Rosenberg2017Minimal}, like the fractal property \cite{Song2007How} and self-similarity property \cite{Song2005Self} of complex networks. These properties have been used in various fields in the network. Currently, lots of relevant studies have been carried out to study the significant properties of the network, like measuring the similarity between nodes to find the same user in different apps \cite{wentao2019similar}; predicting the potential links in networks to find possible relationships in social software \cite{Lu2011Link}; exploring the game theory in networks to find the role of evolutionary game in human progress \cite{wang2017onymity,wang2016statistical,Matja2017Statistical}; measuring the vulnerability of networks to guide the reconstruction of networks \cite{wentao2018evaluating}. In particular, only a part of nodes plays an important role to most network properties, i.e. a small number of individuals has a great influence on society \cite{L2016Vital}. In network, this influence means the propagation, representation, and dynamics of nodes. Thus, finding the influential nodes in networks not only has significant theoretical significance but also practical significance. These nodes would have more important influence to the function and structure of networks \cite{Liao2017Ranking}.

Lots of centrality measures have been proposed to identify these nodes with huge influence in the complex network \cite{Iannelli2018Influencers}, the number of vital nodes is very small, but the impact would be indeed much larger than the other nodes. The classical centrality measures contain Betweenness Centrality \cite{Newman2005A}, Closeness Centrality \cite{Freeman1979Centrality}, Degree Centrality \cite{Newman2003Newman}, PageRank \cite{Brin1998anatomy}, and lots of other measures \cite{wang2018amodified}. In addition, part of the algorithm has been wildly used in various aspects of society, like ranking relevant website \cite{Zhang2019long}, detecting threat and managing disaster \cite{Srinivas2019Community}, designing searching algorithm \cite{Pelusi2018Gravitational,Pelusi2018Neural}; affecting synchronization of interconnected network \cite{Feng2018Synchronization} and so on \cite{jiang2018Correlation,Pelusi2019Redundancy,DNTIJAR2019,Jiang2019Znetwork}. However, these existing centrality have their own limitations. For instance, Betweenness Centrality has a high computational complexity, and lots of nodes' value would be 0 which cannot identify their importance; Closeness Centrality cannot be applied in the network with disconnected components; Degree Centrality considers the neighbor nodes' influence but ignores the influence all over the network. 

Recently, some novel centralities have been proposed in this filed to address this problem. For example, Mariani et al. \cite{Zhou2019Fast} proposed a local centrality measure named social capital can fast identify influencer which is based on the local network structure properties. Andrade et al. \cite{de2019pmean} proposed p-means centrality based on the average of the geodesic distances, and obtained the greatest spreading capacity node in the network. Deng et al. \cite{feiidentifying2018} identified the vital nodes by inverse-square law in the complex network. Zhou et al. \cite{Li2019Identifying} modified the gravity model to detect the influential nodes in the complex network which achieve a good performance. There still are lots of methods used in this filed, such as TOPSIS \cite{Zareie2018TOPSIS}, evidence theory \cite{Liu2019Identifying,MO2019121538evidence}, entropy-based method \cite{Zareie2017Influential}, nodes' relationship \cite{SHEIKHAHMADI2017517online,Zhang2019Groups}, optimal percolation theory \cite{Ferraro2018Finding,Morone2015Influence}, and so on \cite{SHEIKHAHMADI2017517marketing}. 

Because fractal property is important for various fields \cite{Kunze2018Collage,Davide2018Fractal}, it has been applied to compress image \cite{Kunze2017IFSM}, maximize the expected return \cite{LaTorre2018Portfolio}, optimize population size \cite{Davide2019optimal}, give metric between probability distributions \cite{Mendivil2017Computing}, and get solutions of a classical integral equation \cite{KUNZE2019SELF}. The fractal property and self-similarity property in networks can not only show the network's feature \cite{Rosenberg2017Maximal,Rosenberg2017Non}, but also reveal the nodes' properties \cite{wentao2018information}. Recently, Pu et al. \cite{Pu2014Identifying} modified local dimension in the network to identify the influential nodes. Then, Bian et al. \cite{Bian2018Chaos} measured the information dimension of node to rank the influence of node which is a new research perspective. After that, Jiang et al. \cite{wentao2019nodes} proposed the fuzzy local dimension to identify the influential nodes. Thus, the fractal and self-similarity properties have been proved to be significant for nodes' importance identification.

In this paper, a novel centrality measure is proposed based on multi-local dimension which is from the view of the fractal property.This proposed method considers the structure around the central node by the box. The size of box would increase from 1 to the maximum value of the shortest distance from the central node. The information in each box is represented by the number of nodes in the box. Then, a weighting coefficient $q$ is used to deal with the information. Different chosen of $q$ would consider the information in different scale which can cause different representation of multi-local dimension. MLD would degenerate to local information dimension and variant of local dimension when $q = 1$ and $q = 0$ respectively. Finally, the multi-local dimension of node can be obtained by the slope of linear regression. Thus, this proposed measure is a more general model to identify the vital nodes because the existence of coefficient $q$. Some real-world complex networks have been used in this paper, the effectiveness and reasonableness of this proposed method is demonstrated in comparison with some existing centrality measures. Observing from the experiment results, the superiority of this proposed method and the relationship between this proposed method and other comparison methods can be obtained.

The organization of the rest of this paper is as follows. This proposed multi-local dimension is defined in Section \ref{Sec_proposed} to identify the vital spreaders in the network. Meanwhile, some real-world complex networks and existing comparison methods are performed to illustrate the reasonableness and effectiveness of the proposed method in Section \ref{Sec_experiments}. The conclusion is conducted in Section \ref{Conclusion}.

\section{The proposed vital spreaders identification model} \label{Sec_proposed}

\begin{figure}[!htb]
    \centering
    \includegraphics[width=\textwidth]{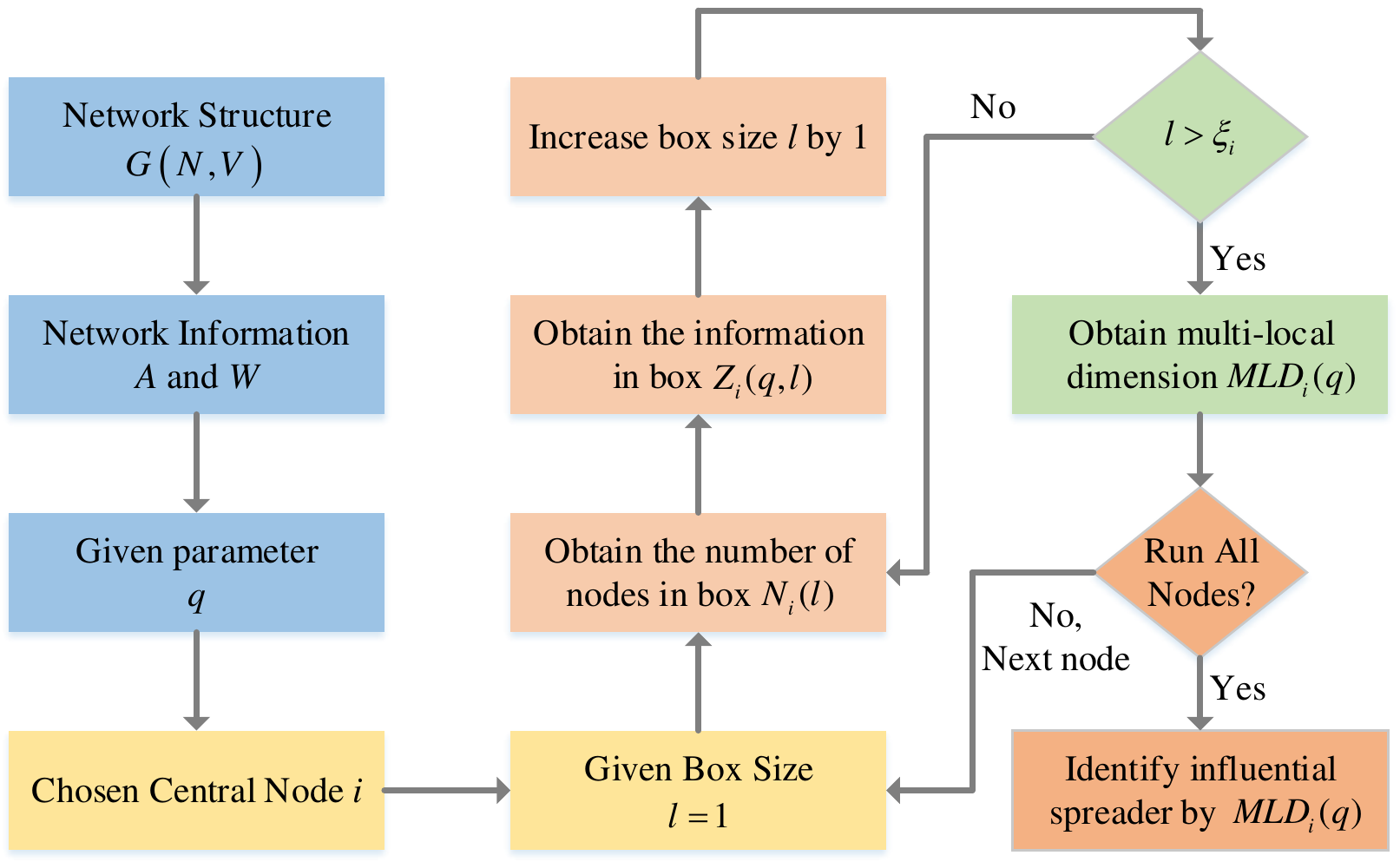}\\
    \caption{\textbf{The flow chart of this proposed method.}}
    \label{fig_flowchart}
\end{figure}

In this section, a novel measure is proposed based on \emph{multi-local dimension} (MLD) to identify the influential spreaders in the complex network. This proposed method can consider the information in boxes with different scale $q$. When $q$ has different values, different expressions of MLD would be given to identify influential nodes. In addition, this proposed method would degenerate to local information dimension and variant of local dimension when $q = 1$ and $q = 0$ respectively. The flow chart of MLD is shown in Fig. \ref{fig_flowchart}.

\subsection{The structure of complex network} \label{Sec_proposed_network}

In a given \emph{complex network} $G\left( {N,E} \right)$, $N$ is the set of nodes and $E$ is the set of edges in the network, ${\left| N \right|}$ and ${\left| E \right|}$ is the number of nodes and edges respectively in the network. Firstly, the \emph{adjacency matrix} $A$ of is given to describe the topological structure of the complex network. The element ${a_{ij}}$ in the adjacency matrix $A$ shows the connection edge between node $i$ and node $j$. ${a_{ij}} = 1$ represents there is an edge between node $i$ and $j$, and ${a_{ij}} = 0$ is the opposite. Then, the \emph{shortest distance} between any two nodes can be obtained by the adjacency matrix $A$ (the known information) through \emph{Dijkstra algorithm} \cite{Dijkstra1959}, and the definition of the shortest distance ${\omega _{ij}}$ between node $i$ and node $j$ can be shown below,
\begin{equation}\label{equ_shortest}
  {\omega _{ij}} = \mathop {\min }\limits_h \left( {{a_{i{h_1}}} + {a_{{h_1}{h_2}}} +  \cdots  + {a_{{h_m}j}}} \right)
\end{equation}
where ${a_{{h_1}{h_2}}}$ is one element of $A$ which can show network's connection, ${h_1},{h_2}, \cdots ,{h_m}$ are the IDs of different nodes. The shortest distance matrix can be constructed by the know shortest distance between any two nodes, and it is denoted as $W$. The element ${\omega _{ij}}$ represents the shortest distance between node $i$ and $j$, and the shortest distance matrix $W$ would be a symmetric matrix. The \emph{maximum value of the shortest distance} from node $i$ is denoted as ${\xi _i}$ and defined as follows,
\begin{equation}\label{equ_max_shortest_distance}
  {\xi _i} = \mathop {\max }\limits_{j \in N,j \ne i} \left( {{\omega _{ij}}} \right)
\end{equation}
where ${\xi _i}$ would vary from the chosen of central node $i$, which can show the scale of locality of central node $i$.

\subsection{The local dimension of complex network} \label{Sec_proposed_local_dimension}

After getting the relevant basic characteristics of complex networks, the local dimension which is the basis of this proposed method is introduced in this section. The \emph{local dimension} (LD) is modified from the fractal dimension and firstly proposed to accurately measure the local property of each node, i.e. the change of dimensionality among vertices in the network \cite{Silva2013Local}. Then, Pu et al. \cite{Pu2014Identifying} modified local dimension to identify the vital nodes in the complex network. In this method, the volume scaling property has been considered in different topological scale. In general, the number of nodes ${B_i}(r)$ within a given radius $r$ (including $r$) for any node $i$ follows a power law which is shown as follows,
\begin{equation}\label{eau_power_law}
  {B_i}(r) \sim {r^{L{D_i}}}
\end{equation}
where ${L{D_i}}$ is the local dimension of node $i$. Thus, the local dimension of any node can be obtained by the slope of double logarithmic curves which is detailed shown below,
\begin{equation}\label{equ_local_dimension_V1}
  L{D_i} = \frac{d}{{d\ln r}}\ln {B_i}(r)
\end{equation}
where $d$ is the symbol of derivative. Due to the discrete property \cite{Ben2004Complex} of complex network, Eq. (\ref{equ_local_dimension_V1}) can be rewritten as follows,
\begin{equation}\label{equ_local_dimension_V2}
  \begin{array}{l}
    L{D_i} = \frac{r}{{{B_i}(r)}}\frac{d}{{dr}}\ln {B_i}(r)\\
    L{D_i} = \frac{r}{{{B_i}(r)}}{b_i}(r)
    \end{array}
\end{equation}
where $r$ is the radius of the box, ${b_i}(r)$ represents the number of nodes whose shortest distance from central node $i$ equal to $r$, and ${{B_i}(r)}$ represents the number of nodes whose shortest distance from central node $i$ is less than or equal to $r$. The radius $r$ whose central node is node $i$ would increase from 1 to ${\xi _i}$, and the local dimension $L{D_i}$ of node $i$ would be the slope of double logarithmic curves.

\subsection{This proposed multi-local dimension of complex network} \label{Sec_proposed_multi_local}

\begin{figure}[!htb]
  \centering
  \includegraphics[width=\textwidth]{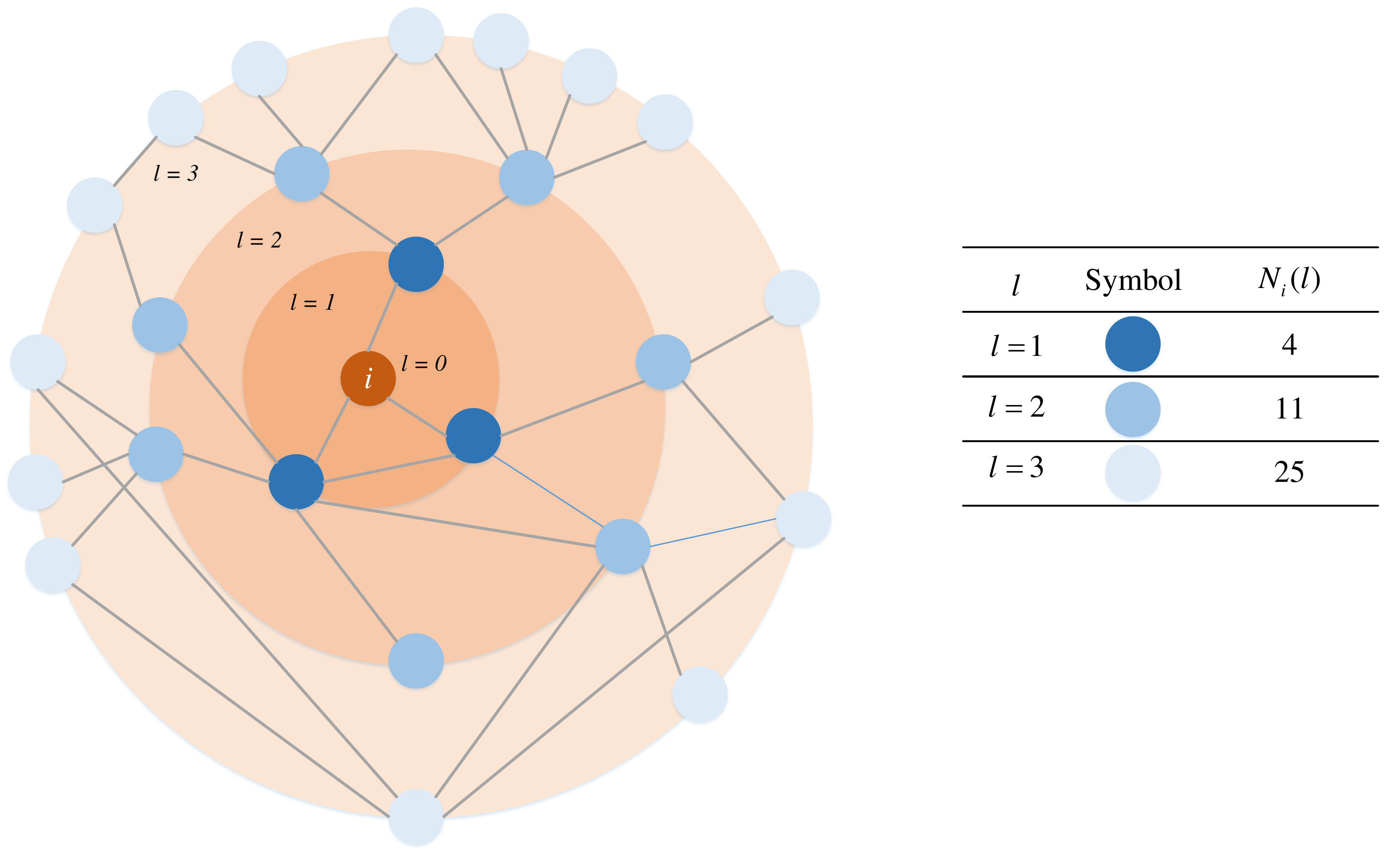}\\
  \caption{\textbf{The example network with different nodes' distance from central node $i$.} These node with different color mean the different shortest distance from central node.}
  \label{fig_example}
\end{figure}

Take node $i$ as the central node as an example in this section. In this proposed method, there is a box covering the network with node $i$ as the central node. The size of the box $l$ would increase from 1 to $\xi _i$, and the entire network would be covered by this box when $l = \xi _i$. These nodes in the network with different distance from central node is shown in Fig. \ref{fig_example}. The information in this box ${\mu _i}(l)$ is related with the number of nodes in this box, and it is defined as follows,

\begin{equation}\label{equ_box_information}
  {\mu _i}(l) = \frac{{{N_i}(l)}}{{\left| N \right|}}
\end{equation}
where ${N_i}(l)$ is the number of nodes covered by this box, i.e. the shortest distance from these nodes to the central node $i$ is less than the size of the box $l$, and ${\left| N \right|}$ is the number of nodes in the network. For this given measure ${\mu _i}(l)$, the partition consideration of the box ${Z_i}(q,l)$ is defined as follows,
\begin{equation}\label{equ_Z_q}
  {Z_i}(q,l) = {\left[ {{\mu _i}(l)} \right]^q}
\end{equation}
where $q$ is the real number ($q \in R$) which can be changed. In addition, $q$ plays a weighting coefficient for Eq. (\ref{equ_Z_q}). In addition, when $q = 0$, the partition consideration ${Z_i}(q,l) = {{{\mu _i}(l)}}$. Thus, the partition consideration of the box would have following property: ${Z_i}(q,l) \ge 0$. 

Then, the multi-local dimension $ML{D_i}(q)$ of node $i$ is defined as follows,
\begin{equation}\label{equ_MLD}
  ML{D_i}(q) = \left\{ {\begin{array}{*{20}{c}}
  {\frac{{{\tau _i}(q,l)}}{{q - 1}},}&{q \ne 1}\\
  {\mathop {\lim }\limits_{l \to 0} \frac{{{Z_i}(1,l)}}{{\ln l}},}&{q = 1}
  \end{array}} \right.
\end{equation}
because the denominator $(q - 1)$ would equal to 0 when $q = 1$, MLD would have different expression in this situation. When $q \ne 1$, the  numerator ${{\tau _i}(q,l)}$ would be defined as follows,
\begin{equation}\label{equ_tau}
  {\tau _i}(q,l) = \mathop {\lim }\limits_{l \to 0} \frac{{\ln {Z_i}(q,l)}}{{\ln l}}
\end{equation}
where ${Z_i}(q,l)$ has been given in (\ref{equ_Z_q}), and $l$ is the size of the box. When $q = 1$, ${{Z_i}(1,l)}$ would be defined as follows,
\begin{equation}\label{equ_Z_q_1}
  {Z_i}(1,l) = {\mu _i}(l)\ln {\mu _i}(l)
\end{equation}
where the partition consideration follows the expression of Shannon entropy.

In this proposed method, the scale of locality for each nodes is different, which is decided by the maximum value of the shortest distance ${\xi _i}$ from central node $i$. The box size $l$ would change from 1 to ${\xi _i}$ for each node. The numerical estimation of MLD would be obtained by the linear regression of $\ln {Z_i}(q,l)/(q - 1)$ against ${\ln l}$ for ${q \ne 1}$, and when $q = 1$, MLD would be obtained by the linear regression of ${{Z_i}(1,l)}$ against ${\ln l}$. Similar to multi-fractal dimension, the multi-local dimension can degenerate to other dimensions with different values of $q$, and it is detailed shown below.

\begin{itemize}
  \item When $q = 1$, MLD would degenerate to local information dimension \cite{Wen2019local}.
  \item When $q = 0$, MLD would degenerate to variant of local dimension \cite{Pu2014Identifying}.
\end{itemize} 

Both of these two measures have been applied to identify the influential spreaders in the complex network. Thus, this proposed method MLD is a more general method.

\subsection{Vital spreaders identification} \label{Sec_proposed_identification}

When the multi-local dimension $ML{D_i}$ is obtained, the importance of spreaders can be ranked by the value of multi-local dimension. Different with previous methods, the spreader would be more important with smaller MLD. The details can be shown in Section \ref{Sec_experiments}.

\section{Experimental study} \label{Sec_experiments}

In this section, four different scale real-world complex networks and three comparison methods are used in this section to show the reasonableness and effectiveness of this proposed method. Four kinds of experiments are utilized in this section, including giving top-10 nodes lists, obtaining the individuation of each nodes' rank results, measuring the infectious ability of initial nodes, describing the relationship between different measures and infectious ability obtained by SI model.

\subsection{Data} \label{Sec_exp_data}

There are four different scale real-world complex networks used in this section to show the effectiveness and reasonableness of this proposed method, and they are:
\begin{enumerate}[1)]
  \item The Zacharys Karate network: This network demonstrates the relationship between many individuals in one USA university karate club; 
  \item The Jazz musicians network: This network shows the collaborations between different jazz musicians;
  \item The USA airline network: This network represents the airlines between the big city airports in the USA;
  \item The Political blogs network: This network demonstrates the blogs' connection in two camps in the USA.
  \end{enumerate}
These network can download from \url{http://vlado.fmf.uni-lj.si/pub/networks/data/}. The detailed structural information of these four networks are shown in Table \ref{table_network_property}. ${\left| N \right|}$ and ${\left| E \right|}$ is the number of nodes and edges in the network respectively. $\left\langle k \right\rangle $ and ${k_{\max }}$ is the average value and maximum value of degree of node in the entire network. $\left\langle \omega \right\rangle $ and ${\omega_{\max }}$ represents the average value and maximum value of the shortest distance in the network. 

\begin{table}[!htb]
\centering
\caption{\textbf{The topological properties of real-world complex networks.}}
\begin{tabular}{ccccccc}
\hline
\hline
Network & ${\left| N \right|}$ & ${\left| E \right|}$ & $\left\langle k \right\rangle $ & ${k_{\max }}$ & $\left\langle \omega \right\rangle $ & ${\omega_{\max }}$ \\
\hline
Karate            & 34   & 78    & 4.5882  & 17  & 2.4082 & 5 \\
Jazz              & 198  & 5484  & 27.6970 & 100 & 2.2350 & 6 \\
USAir             & 332  & 2126  & 12.8072 & 139 & 2.7381 & 6 \\
Political blogs   & 1222 & 19021 & 27.3552 & 351 & 2.7375 & 8 \\
\hline
\hline
\end{tabular}
\label{table_network_property}
\end{table}

\subsection{Existing centrality measures} \label{Sec_exp_measures}

Before the experiment begins, let's introduce some existing centrality measures to identify the influential nodes as comparison methods in this section. Because MLD would degenerate to local information dimension and variant of local dimension, these two measures would not be used as comparison measures in this section.

\begin{definition}\label{defi_BC}
  Betweenness centrality (BC) \cite{Newman2005A}. The betweenness centrality of node $i$ is expressed as ${C_B}(i)$, and it is defined as follows,
  \begin{equation}\label{equ_BC}
    {C_B}(i) = \sum\limits_{s,t \ne i} {\frac{{{g_{st}}(i)}}{{{g_{st}}}}}
  \end{equation}
  where ${{g_{st}}(i)}$ means the shortest path between node $s$ and node $t$ which go through node $i$, and ${{g_{st}}}$ means the shortest path between node $s$ and node $t$. Node $s$ and node $t$ would traverse all nodes in the network. Thus, BC highlights the intermediary role of selected node.
\end{definition}

\begin{definition}\label{defi_CC}
  Closeness centrality (CC) \cite{Freeman1979Centrality}. The closeness centrality of node $i$ is expressed as ${C_C}(i)$, and it is defined as follows,
  \begin{equation}\label{equ_CC}
    {C_C}(i) = {\left( {\sum\limits_{j \in N} {{\omega _{ij}}} } \right)^{ - 1}}
  \end{equation}
  where ${{\omega _{ij}}}$ is the shortest distance between node $i$ and node $j$ which belongs to the shortest distance matrix $W$, and node $j$ would traverse all nodes in the network. Thus, CC highlights that the selected node can quickly reach any node in the network.
\end{definition}

\begin{definition}\label{defi_DC}
  Degree centrality (DC) \cite{Newman2003Newman}. The degree centrality of node $i$ is expressed as ${C_D}(i)$, and it is defined as follows,
  \begin{equation}\label{equ_DC}
    {C_D}(i) = \sum\limits_{j \in N} {{a_{ij}}} 
  \end{equation}
  where ${{a_{ij}}}$ is the element in adjacency matrix $A$, and node $j$ would traverse all nodes in the network. When there is an edge between node $i$ and node $j$, ${{a_{ij}}}$ would equal to 1, and ${{a_{ij}} = 0}$ represents the opposite situation. In fact, the degree centrality of node $i$ represents the number of edges connected with node $i$. Thus, DC highlights the number of neighbor nodes around selected node in the network.
\end{definition}

\subsection{Experiment I: Top-10 nodes} \label{Sec_exp_top}

In this experiment, the top-10 nodes lists are obtained by different measures to show the difference and correlation between these methods, and these lists are shown in Table \ref{table_top_10}. Because these methods consider different parts of information in the network, their rank lists may be different with the others. When two methods' top-10 nodes lists are similar, their consideration information would be similar. In addition, the same nodes between MLD and other methods can bring more credibility to this proposed method. These nodes which only appear in MLD result would have a significant improvement to the propagation process. 

\begin{enumerate}[1)]
  \item Observing the result in Karate network from Table \ref{table_top_10}, the most similar lists to MLD is BC,and there are eight same nodes between BC and MLD. The number of same nodes between CC, DC, and MLD is five and six nodes respectively, which is relatively low compared to the results between BC and MLD. The result means that the most similar method to MLD is BC, which is different from the later experiments. 
  \item In Jazz network, the result between BC and MLD is the most dissimilar, and there are only three same nodes between these two measures which is the lowest same number of all results. Compared CC with this proposed method, there are 7 same top-10 nodes. In addition, the top-10 nodes lists are almost the same using MLD and DC, and it is 9 same nodes in the top-10 nodes lists. 
  \item Similar to Jazz network's result, the number of the same top-10 nodes between DC and MLD in USAir network is the highest in three comparison methods, and it is 8 same nods. There is six same nodes between this proposed method and CC in this top-10 nodes lists. The number of same top-10 nodes between BC and MLD is also the lowest in three comparison methods, and there is only four same nodes between two measures, which means there are difference between BC and MLD. The most influential node identified by three comparison methods and MLD is the same, and it is node 118, which means the accuracy of this proposed method.
  \item Observing the Political blogs network's result from Table \ref{table_top_10}, all comparison methods have many same nodes in top-10 lists. CC and DC both have nine same top-10 nodes with MLD, and this only one different node is the ninth and tenth node respectively. The lowest number of same nodes is between BC and MLD, and it is seven, which is bigger than the results in other networks. The top-2 nodes are the same in CC, DC, and MLD. From the result in this network, it can be found the similarity between this proposed method and other comparison methods is high. 
\end{enumerate}

In conclusion, observing from the number of the same top-10 nodes, this proposed method has close performance with DC, and it is far from BC. The effectiveness and superiority would be demonstrated in the following sections. Because this proposed method MLD can degenerate to local information dimension and variant of local dimension, these two measures would not be contained in the following experiments.

\begin{table}[!htb]
        \centering
    \caption{\textbf{The top-10 nodes ranked by different centrality methods in six real-world complex networks.}}
    \begin{tabular}{c|cccc|cccc}
    \hline
    \hline
    \multirow{2}{*}{Rank}  & \multicolumn{4}{c|}{Karate Network} & \multicolumn{4}{c}{Jazz Network} \\ \cline{2-9} 
                           & BC   & CC   & DC  & MLD & BC   & CC   & DC  & MLD   \\ \cline{1-9} 
    1                      & 1    & 1    & 34  & 34  & 136  & 136  & 136 & 60    \\
    2                      & 3    & 3    & 1   & 1   & 60   & 60   & 60  & 136   \\
    3                      & 34   & 34   & 33  & 33  & 153  & 168  & 132 & 132   \\
    4                      & 33   & 32   & 3   & 24  & 5    & 70   & 168 & 83    \\
    5                      & 32   & 33   & 2   & 3   & 149  & 83   & 70  & 168   \\
    6                      & 6    & 14   & 32  & 2   & 189  & 132  & 108 & 99    \\
    7                      & 2    & 9    & 4   & 30  & 167  & 194  & 99  & 108   \\
    8                      & 28   & 20   & 24  & 6   & 96   & 122  & 158 & 158   \\
    9                      & 24   & 2    & 14  & 7   & 115  & 174  & 83  & 194   \\
    10                     & 9    & 4    & 9   & 28  & 83   & 158  & 7   & 7     \\ \hline
    \multirow{2}{*}{Rank}  & \multicolumn{4}{c|}{USAir Network}  & \multicolumn{4}{c}{Political blogs Network} \\ \cline{2-9} 
                           & BC   & CC   & DC  & MLD &  BC  & CC   & DC  & MLD   \\ \cline{1-9} 
    1                      & 118  & 118  & 118 & 118 & 12   & 28   & 12  & 12    \\
    2                      & 8    & 261  & 261 & 261 & 304  & 12   & 28  & 28    \\
    3                      & 261  & 67   & 255 & 152 &  94  & 16   & 304 & 304   \\
    4                      & 47   & 255  & 182 & 230 & 28   & 14   & 14  & 14    \\
    5                      & 201  & 201  & 152 & 255 & 145  & 36   & 16  & 16    \\
    6                      & 67   & 182  & 230 & 182 & 6    & 67   & 94  & 94    \\
    7                      & 313  & 47   & 166 & 112 & 16   & 94   & 6   & 6     \\
    8                      & 13   & 248  & 67  & 147 & 300  & 35   & 67  & 67    \\
    9                      & 182  & 166  & 112 & 166 & 163  & 145  & 35  & 35    \\
    10                     & 152  & 112  & 201 & 293 & 35   & 304  & 145 & 36    \\
    \hline
    \hline
    \end{tabular}
    \label{table_top_10}
    \end{table}

\subsection{Experiment II: Individuation} \label{Sec_exp_Individuation}

Then, different methods' capability to identify influential nodes are explored in this section. The importance of these nodes with same score (frequency) cannot be distinguished correctly, but it is a common situation in this field. Thus, a more useful method should be found to give nodes as individual values as possible. If one method can give all these nodes with unique score, this method can give a reasonable importance rank lists to avoid ambiguous rank results. So the individual of method can be considered an effectiveness indicator to show the quantity of different methods. The higher the individual of one method is, the more effective this method is. 

\begin{figure}[!htb]
  \subfigure[Karate network]{
  \label{fig_Fre_Karate} 
  \begin{minipage}[b]{0.5\textwidth}
  \centering
  \includegraphics[width=\textwidth]{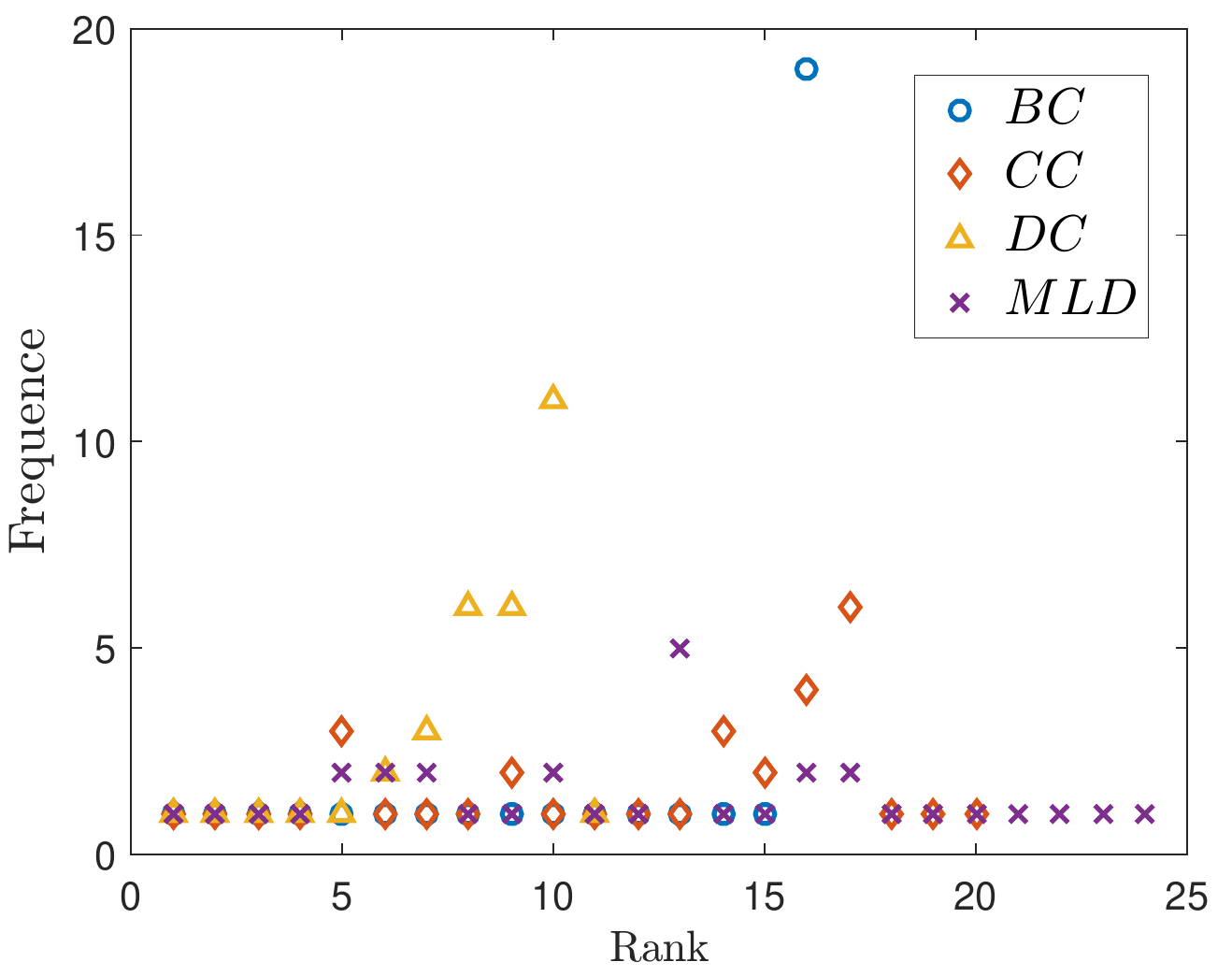}
  \end{minipage}}%
  \subfigure[Jazz network]{
  \label{fig_Fre_Jazz} 
  \begin{minipage}[b]{0.5\textwidth}
  \centering
  \includegraphics[width=\textwidth]{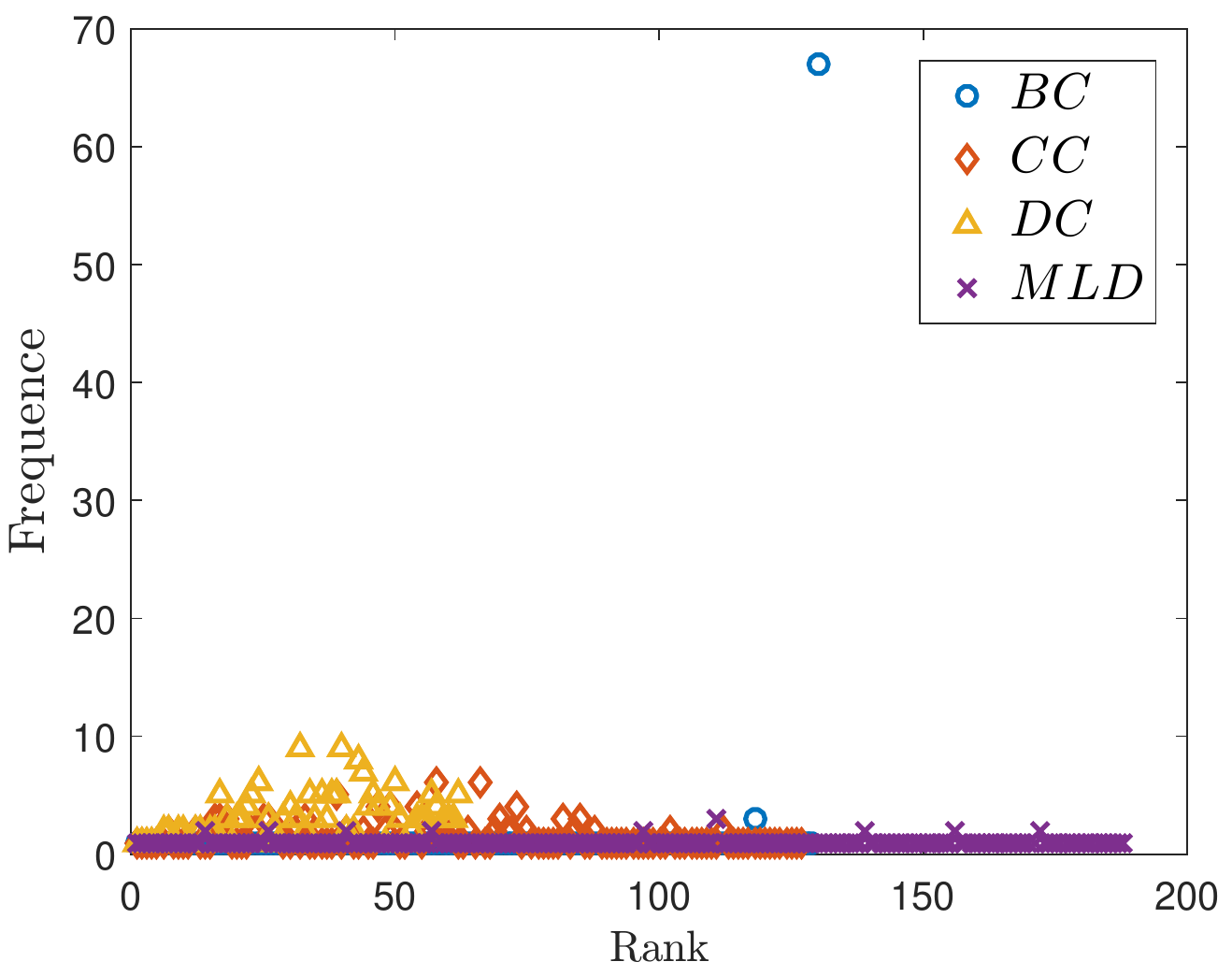}
  \end{minipage}}

  \subfigure[USAir network]{
    \label{fig_Fre_USAir} 
    \begin{minipage}[b]{0.5\textwidth}
    \centering
    \includegraphics[width=\textwidth]{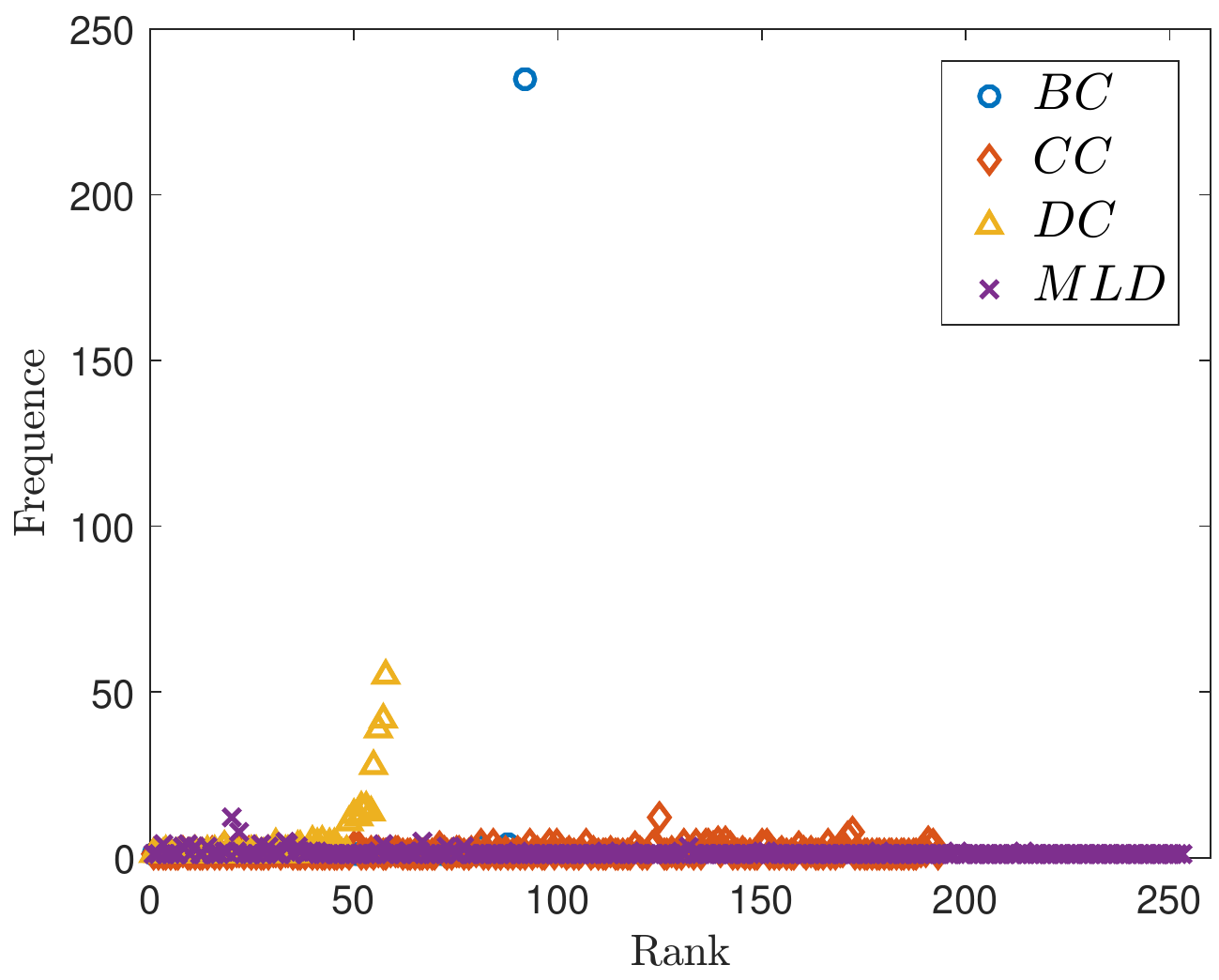}
    \end{minipage}}%
    \subfigure[Political blogs network]{
    \label{fig_Fre_Poli} 
    \begin{minipage}[b]{0.5\textwidth}
    \centering
    \includegraphics[width=\textwidth]{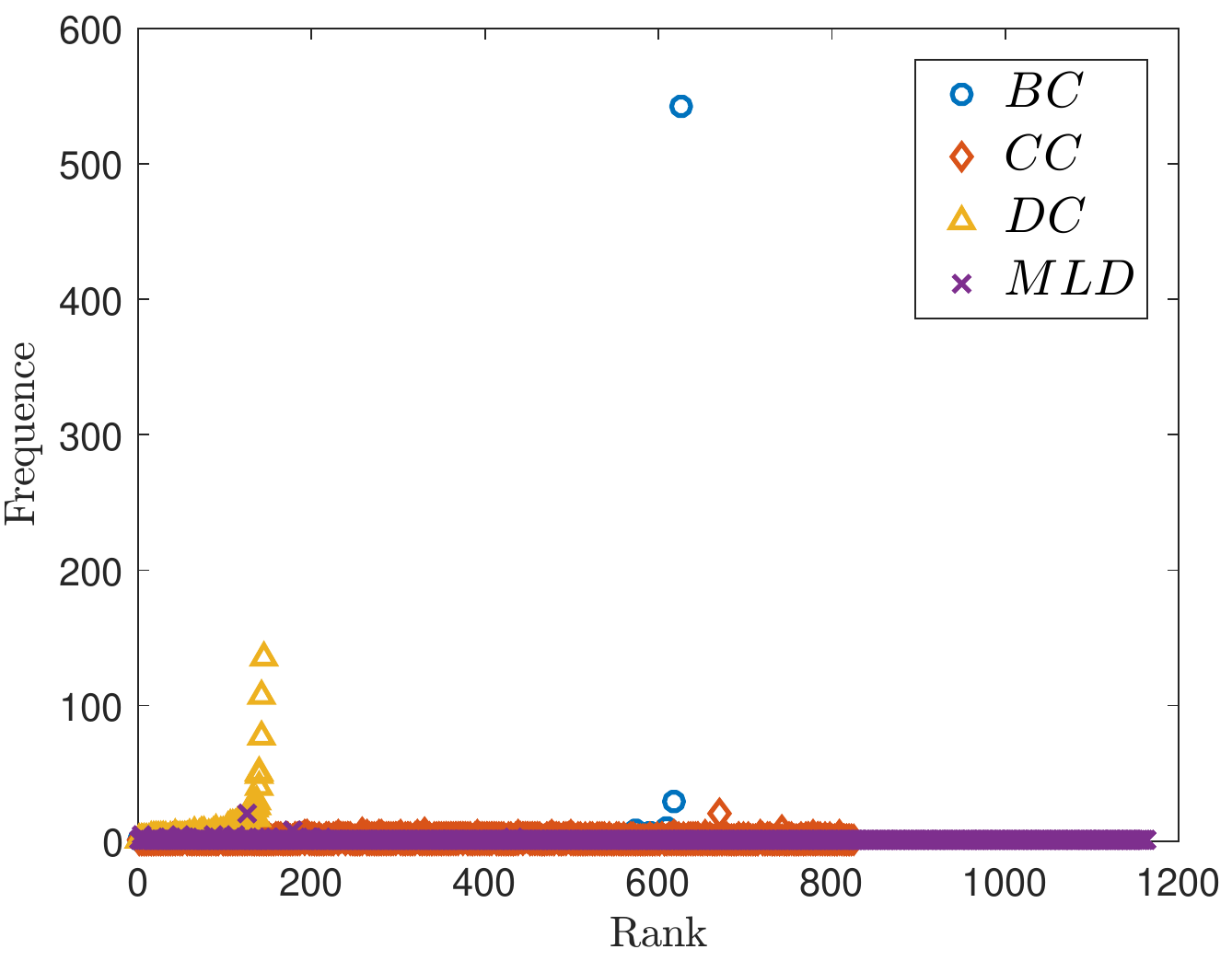}
    \end{minipage}}
  \caption{\textbf{The frequency of nodes in each rank obtained by different measures in real-world complex networks.} Less nodes with same score and more rankings mean the effectiveness of this method. It can be found MLD is the most effectiveness method in these four networks.}
  \label{fig_Fre} 
  \end{figure}

\begin{definition}
The individuation of one method is defined as follows,
\begin{equation}\label{equ_individuation}
  \gamma ( \cdot ){\rm{ = }}\frac{{{N_S}( \cdot )}}{{\left| N \right|}}
\end{equation}
where ${{N_S}( \cdot )}$ is the number of nodes with unique score, ${\left| N \right|}$ is the number of nodes in the entire network. $\gamma ( \cdot )$ is the individuation of one method. 
\end{definition}

The frequency of nodes in each rank obtained by different measures is shown in Fig. \ref{fig_Fre}. In these four network, it can be found that MLD has the least number of nodes in the same rank, and there are more ranks in this proposed method. In contrast, other three comparison methods have more nodes with same rank. In these four networks, DC has the least ranks which means there are lots of nodes with the same ranks. The frequency of nodes in most of the top ranks is relatively low in BC, but the last few ranks have a very high frequency (almost half of nodes), which means BC cannot identify these nodes with low $C_B$. CC can give a relatively reasonable ranks, because most of the frequency in each ranks is low and there are relatively more ranks. However, compared with CC, MLD is more effective to identify the influential nodes. That is because the frequency of nodes in each rank is the least in these four methods, and there are the most ranks (almost one node have one unique ranks) in these networks.

\begin{table}[!htb]
  \centering
  \caption{\textbf{The individuation $\gamma ( \cdot )$ of different methods in real-world complex networks.}}
  \begin{tabular}{ccccc}
  \hline
  \hline
  Network & $\gamma (BC)$ & $\gamma (CC)$ & $\gamma (DC)$ & $\gamma (MLD)$  \\
  \hline
  Karate            & 0.4705 & 0.5882 & 0.3235 & \textbf{0.7058}  \\
  Jazz              & 0.6565 & 0.6414 & 0.3131 & \textbf{0.9494} \\
  USAir             & 0.2771 & 0.5813 & 0.1746 & \textbf{0.7620} \\
  Political blogs   & 0.5114 & 0.6743 & 0.1178 & \textbf{0.9525} \\
  \hline
  \hline
  \end{tabular}
  \label{table_individual}
  \end{table}

The individuation $\gamma ( \cdot )$ of different methods in real-world complex networks are shown in Table \ref{table_individual}, where the highest $\gamma ( \cdot )$ is bold. It can be found that MLD have the highest individuation $\gamma ( \cdot )$ in these four methods, and DC has the lowest individuation $\gamma ( \cdot )$. These results mean that this proposed method is an effective method to identify the influential nodes in the complex network.

\subsection{Experiment III: SI model} \label{Sec_exp_SI}

In this section, Susceptible-Infected (SI) model \cite{L2016Vital} is applied to show the effectiveness and reasonableness of this proposed method. The details of SI model is introduced below. 
\begin{enumerate}[Step 1]
  \item For the entire network, all nodes are classified into two states, and they are \emph{susceptible state} and \emph{infected state}.
  \item At the beginning, the top-10 nodes obtained by centrality measure (shown in Table \ref{table_top_10}) are set as infected state, and the other nods are set as susceptible state. 
  \item When the infection process begins, these susceptible node can be affected by their neighbor nodes with a given probability (spreading ability) $\lambda = (1/2)^\beta$ in each time $t$. In addition, the total number of susceptible nodes and infectious nodes equals to ${\left| N \right|}$ in any time $t$.
  \item Once the susceptible node is infected into infectious node, it cannot return to the susceptible state, i.e. it is the irreversible process.
  \item The number of infectious nodes $F(t)$ would continue to increase over time $t$ until all nodes are infected.
\end{enumerate}

\begin{figure}[!htb]
  \subfigure[Karate network]{
  \label{fig_SI_Karate} 
  \begin{minipage}[b]{0.5\textwidth}
  \centering
  \includegraphics[width=\textwidth]{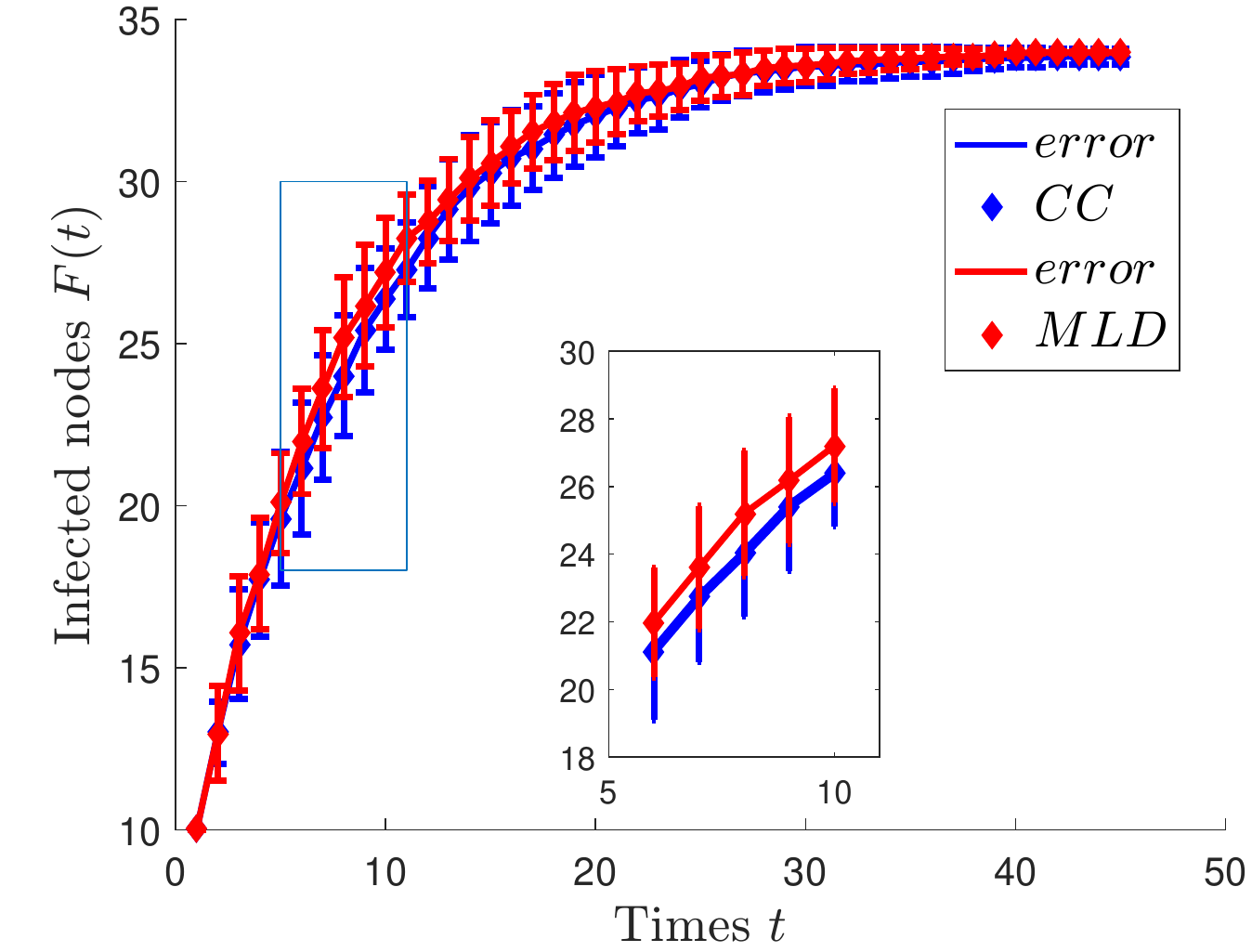}
  \end{minipage}}%
  \subfigure[Jazz network]{
  \label{fig_SI_Jazz} 
  \begin{minipage}[b]{0.5\textwidth}
  \centering
  \includegraphics[width=\textwidth]{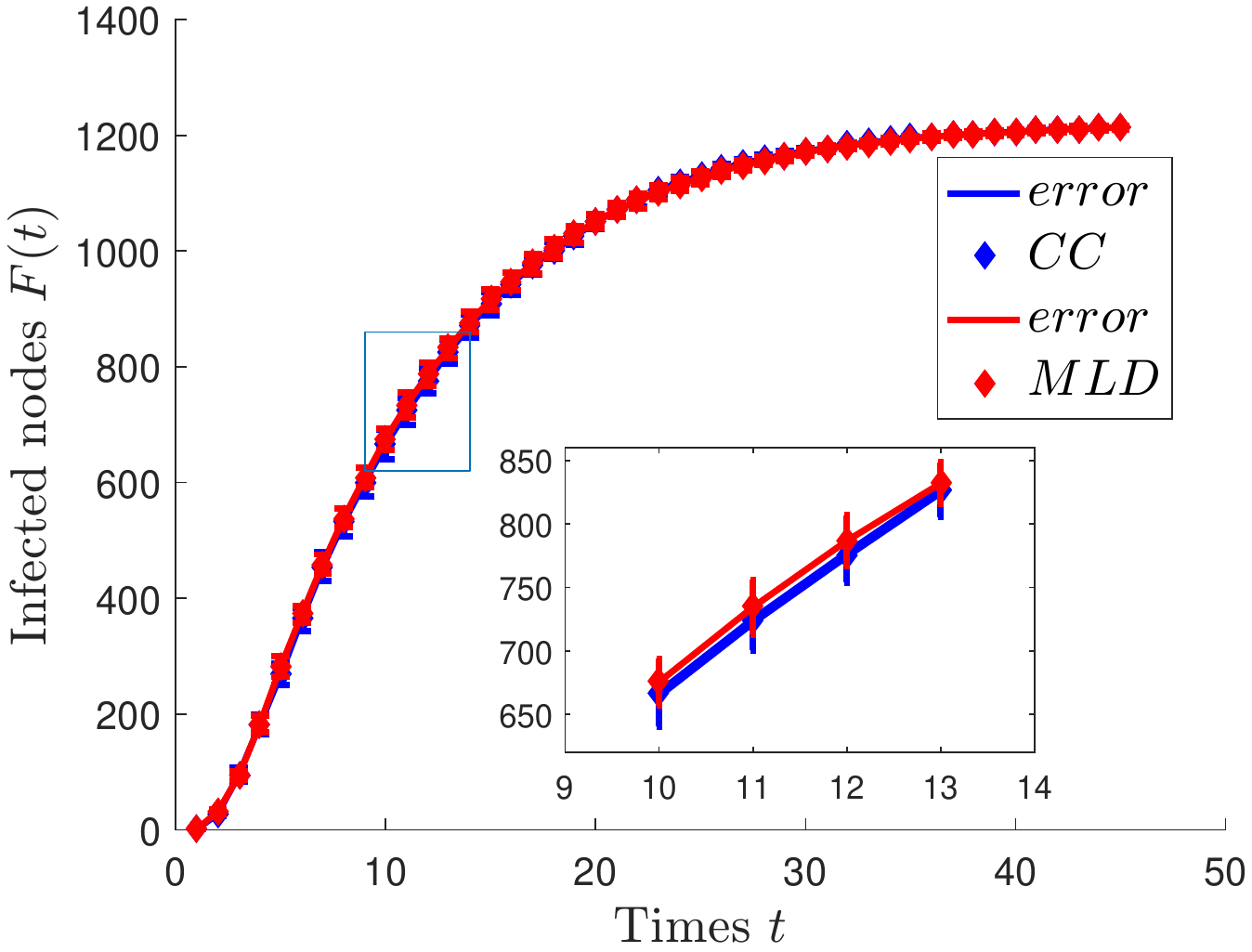}
  \end{minipage}}

  \subfigure[USAir network]{
    \label{fig_SI_USAir} 
    \begin{minipage}[b]{0.5\textwidth}
    \centering
    \includegraphics[width=\textwidth]{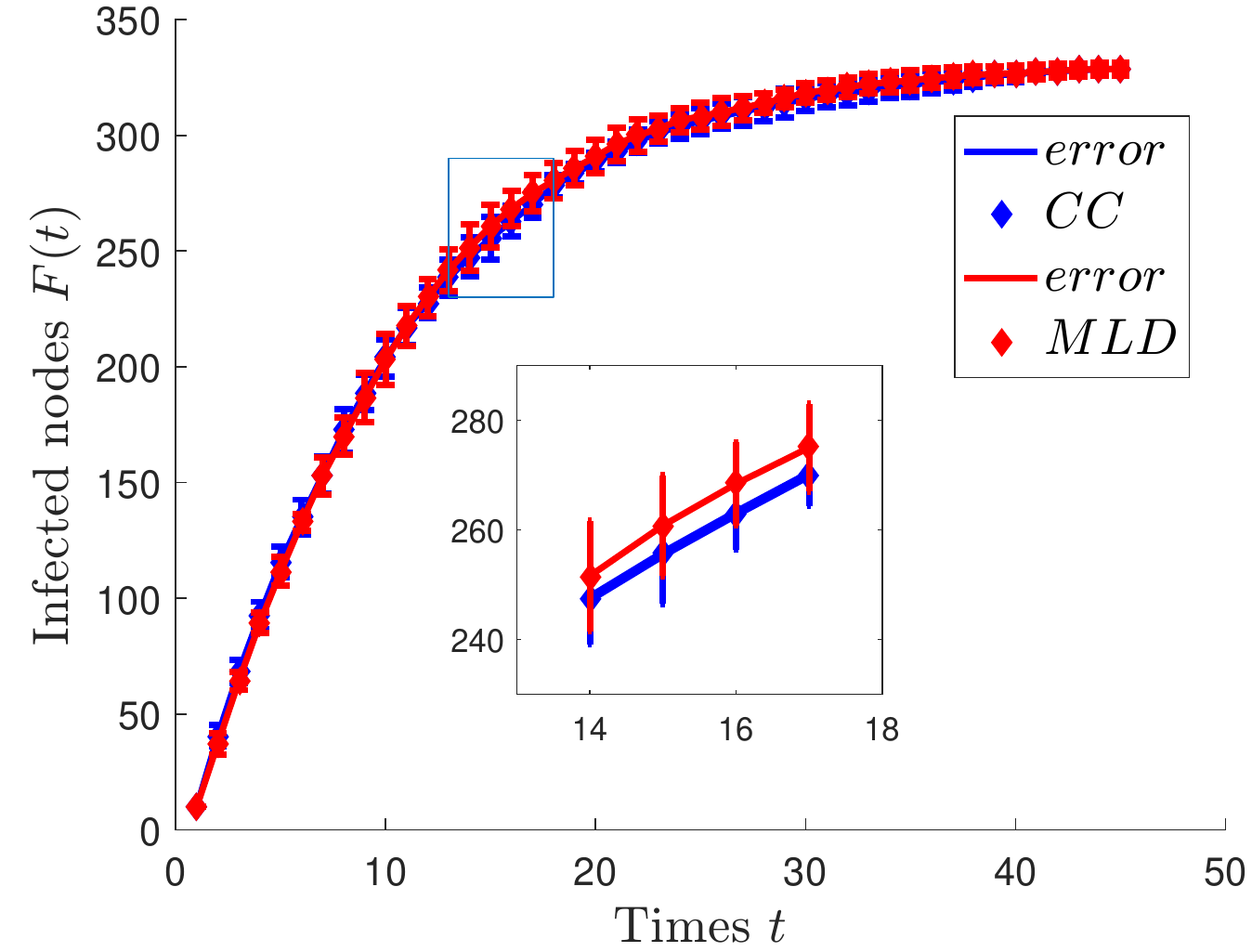}
    \end{minipage}}%
    \subfigure[Political blogs network]{
    \label{fig_SI_Poli} 
    \begin{minipage}[b]{0.5\textwidth}
    \centering
    \includegraphics[width=\textwidth]{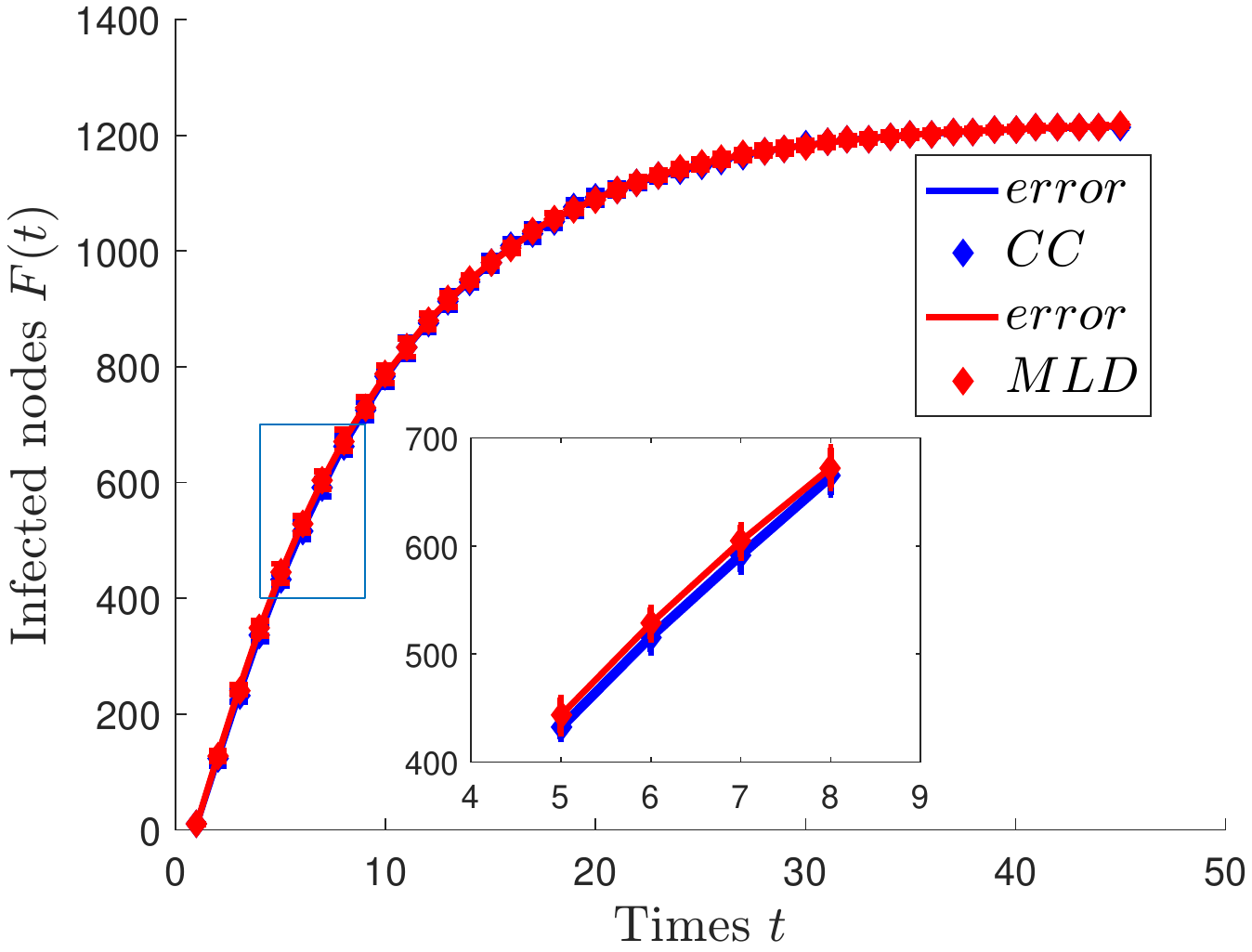}
    \end{minipage}}
  \caption{\textbf{The number of infectious nodes $F(t)$ with different initial nodes obtained by CC and MLD in real-world complex networks.} The details of these figure are enlarged to facilitate observation. The high $F(t)$ in each time $t$ means the strong infectious ability of these initial nodes.}
  \label{fig_SI} 
  \end{figure}


These initial nodes with higher infection ability would infect the entire network as early as possible, so the number of infectious node $F(t)$ can be a effective indicator to show the infection ability of initial nodes, i.e. the importance of initial nodes. More infectious nodes in each time $t$ is, higher infectious ability these initial nodes are, more important these initial nodes are. Because CC considers the nodes' distance from the selected node, which is similar to this proposed method, CC is selected as the comparison method in this section. In these networks, all results $F(t)$ would average the results of 30 SI experiments with $\beta = 3$, and the results are shown in Fig. \ref{fig_SI}.

Observing from Fig. \ref{fig_SI}, the number of infectious nodes $F(t)$ continue to increase over all time $t$. In Karate network shown in Fig. \ref{fig_SI_Karate}, the infection ability of initial nodes obtained by MLD is clearly superior to CC, and it can be seen that $F(t)$ obtained by MLD is larger than $F(t)$ obtained by CC from the whole process. In Jazz network shown in Fig. \ref{fig_SI_Jazz}, the performance of MLD is better than CC which can be seen from early and middle propagation process in SI model. In USAir network shown in Fig. \ref{fig_SI_USAir}, the infection ability of initial nodes obtained by MLD is superior than these nodes obtained by CC, and it can be seen from the middle and late term of SI model, $F(t)$ obtained by MLD is larger than $F(t)$ obtained by CC in this term. In Political blogs network shown in Fig. \ref{fig_SI_Poli}, MLD is slightly better than CC, because they are almost same in the propagation process. But the number of infectious nodes $F(t)$ obtained by MLD is bigger than the number obtained by CC between 5 to 15 time. In conclusion, this proposed method have a superiority performance in most of experiments, and some times the performance of MLD is close to the comparison method.

\subsection{Experiment IV: The relationship between different methods} \label{Sec_exp_relationship}

\begin{figure}[!htb]
  \subfigure[Karate network]{
  \label{fig_Karate_CC} 
  \begin{minipage}[b]{0.5\textwidth}
  \centering
  \includegraphics[width=\textwidth]{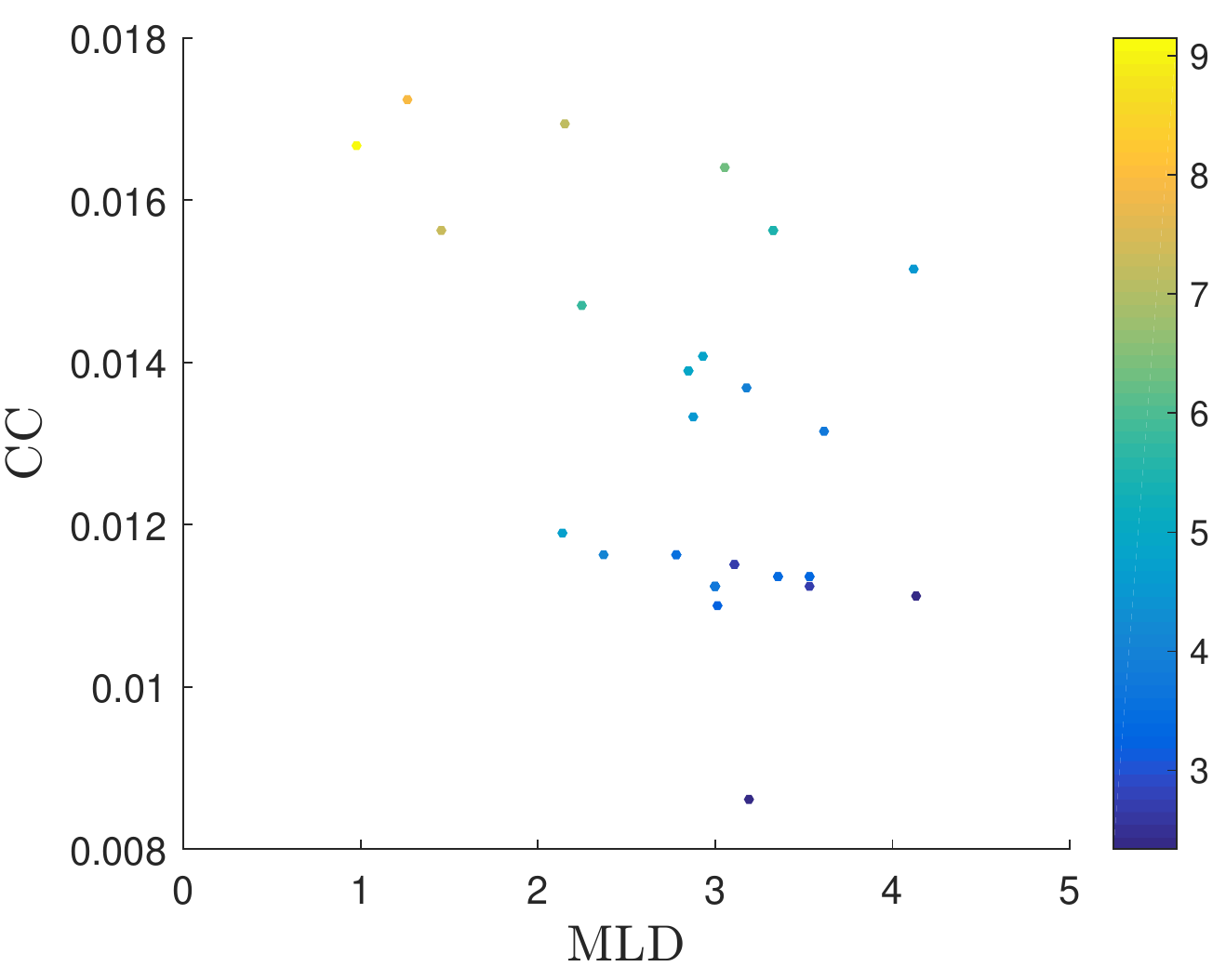}
  \end{minipage}}%
  \subfigure[Jazz network]{
  \label{fig_Jazz_CC} 
  \begin{minipage}[b]{0.5\textwidth}
  \centering
  \includegraphics[width=\textwidth]{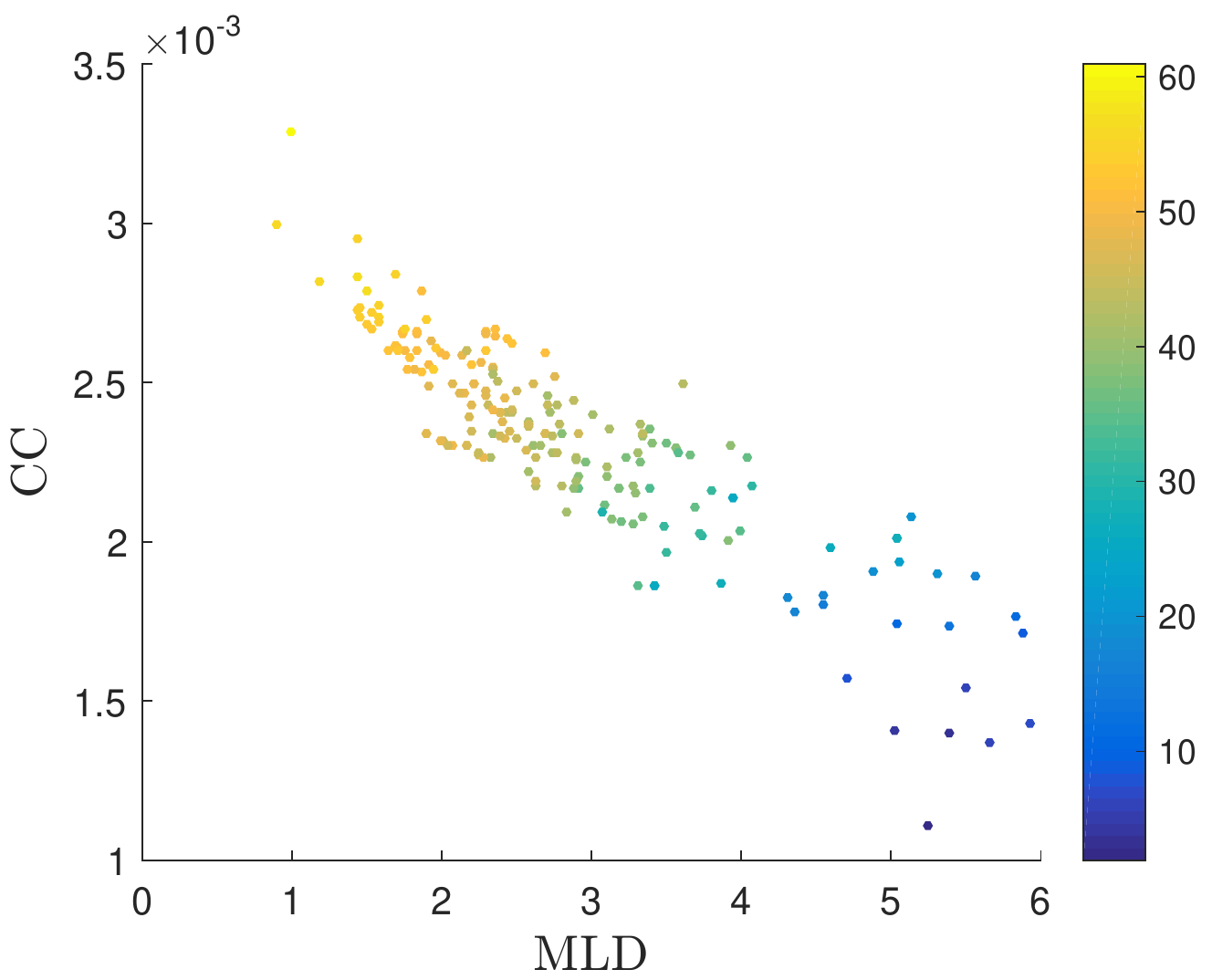}
  \end{minipage}}

  \subfigure[USAir network]{
    \label{fig_USAir_CC} 
    \begin{minipage}[b]{0.5\textwidth}
    \centering
    \includegraphics[width=\textwidth]{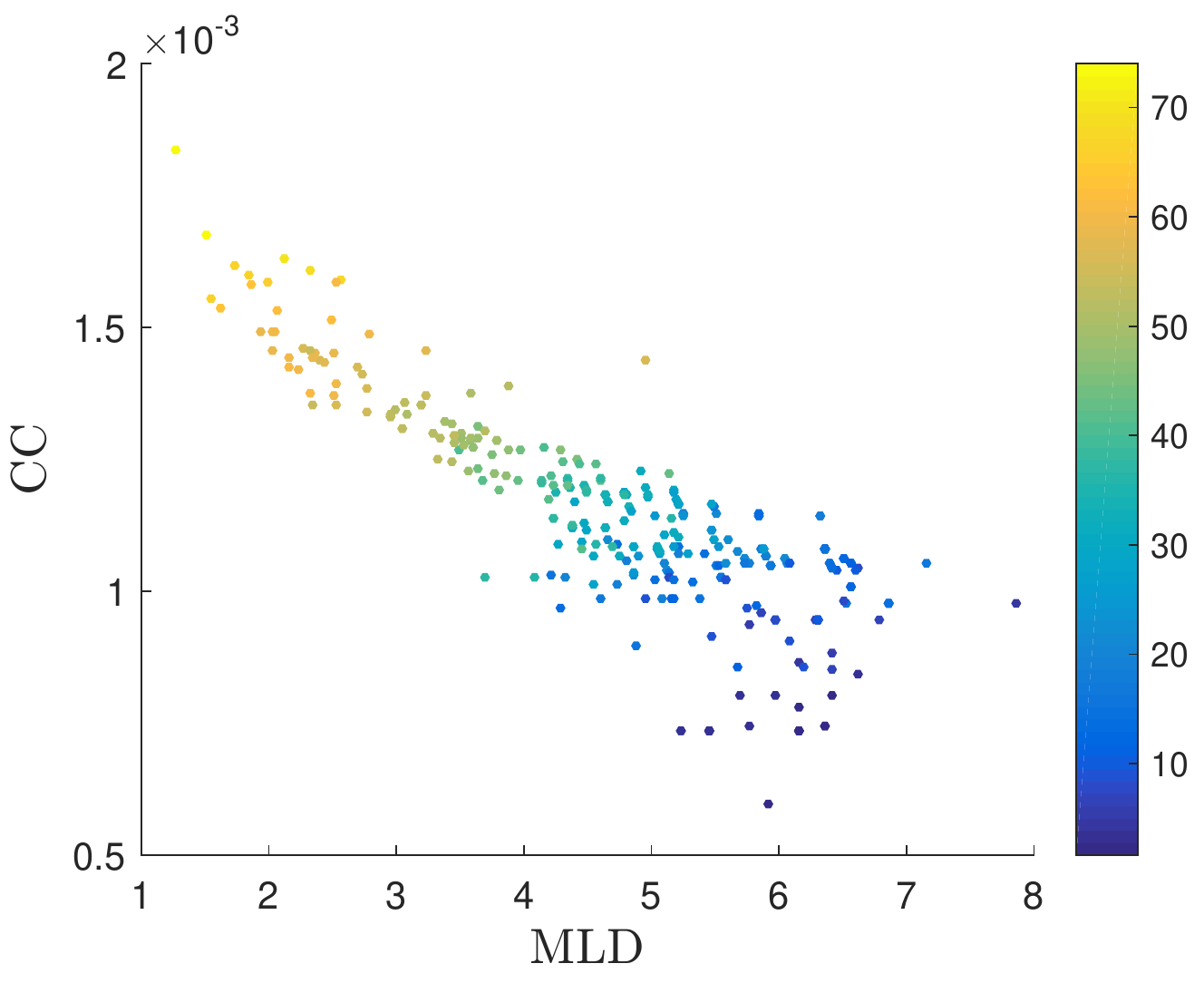}
    \end{minipage}}%
    \subfigure[Political blogs network]{
    \label{fig_Poli_CC} 
    \begin{minipage}[b]{0.5\textwidth}
    \centering
    \includegraphics[width=\textwidth]{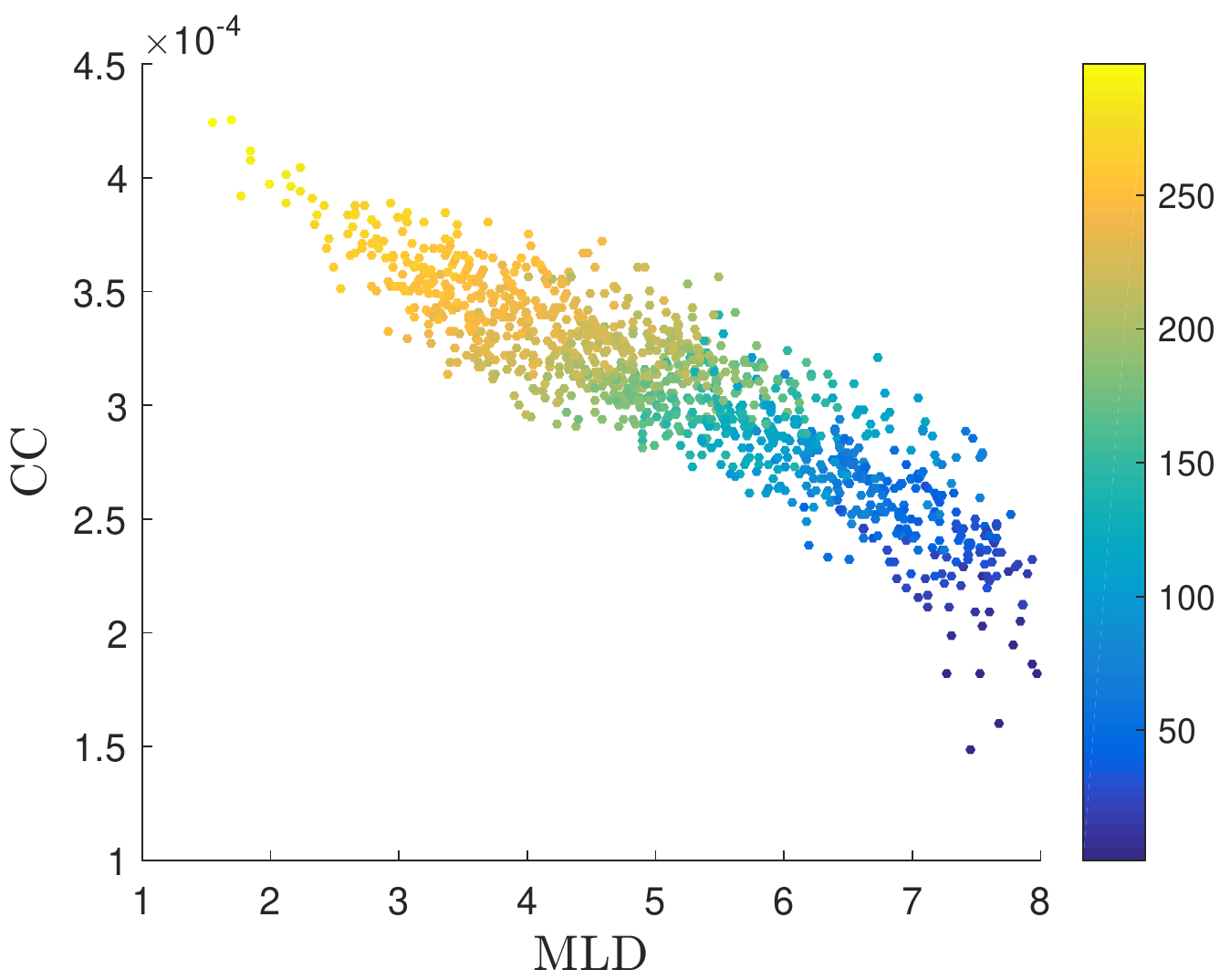}
    \end{minipage}}
  \caption{\textbf{The relationship between MLD and CC when $\lambda = 0.05$ in real-world networks.} The value on the horizontal and vertical axes means the value obtained by MLD and CC respectively, and the color of point means the infectious ability obtained by SI model.}
  \label{fig_COrre_CC} 
  \end{figure}

  \begin{figure}[!htb]
    \subfigure[Karate network]{
    \label{fig_Karate_DC} 
    \begin{minipage}[b]{0.5\textwidth}
    \centering
    \includegraphics[width=\textwidth]{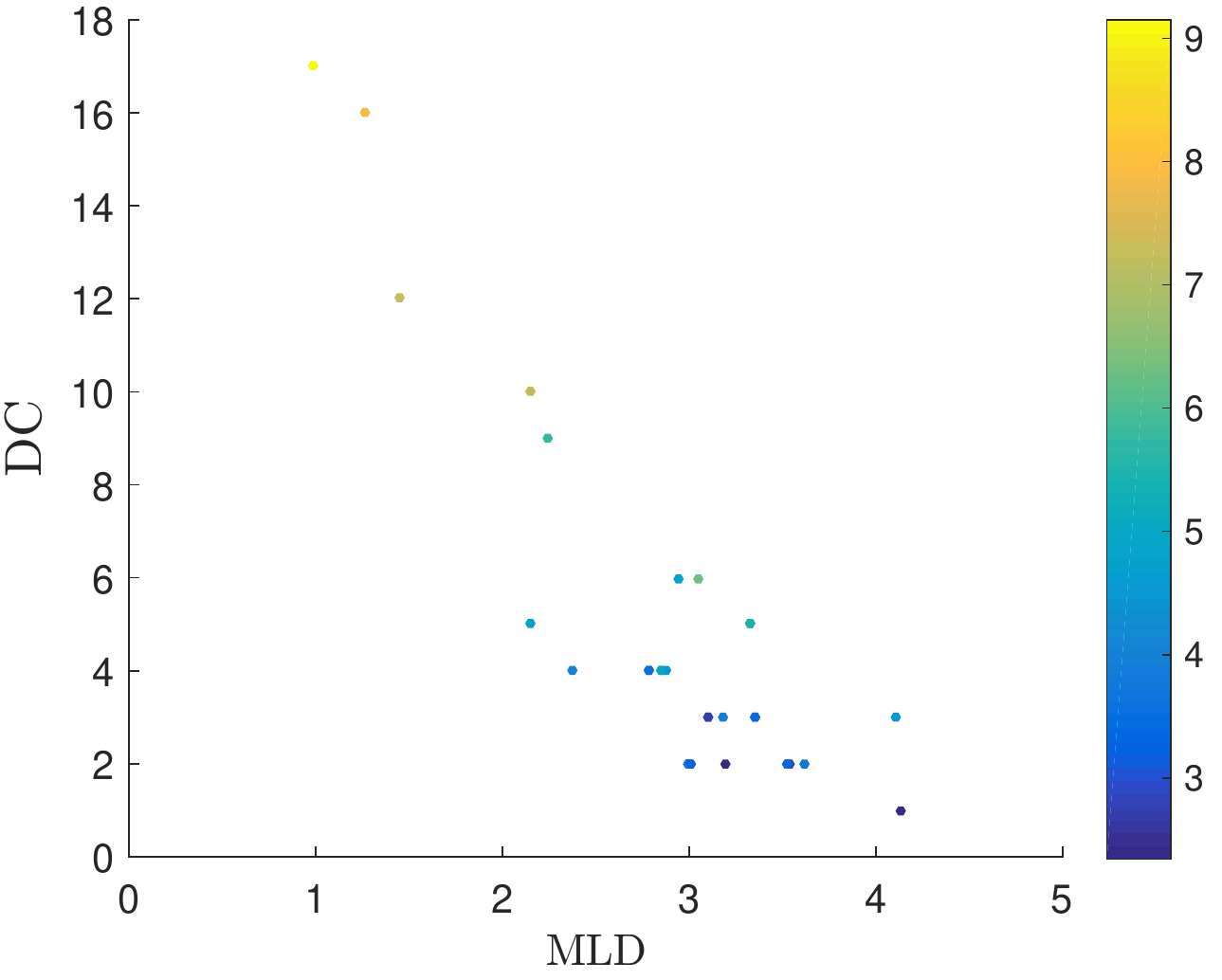}
    \end{minipage}}%
    \subfigure[Jazz network]{
    \label{fig_Jazz_DC} 
    \begin{minipage}[b]{0.5\textwidth}
    \centering
    \includegraphics[width=\textwidth]{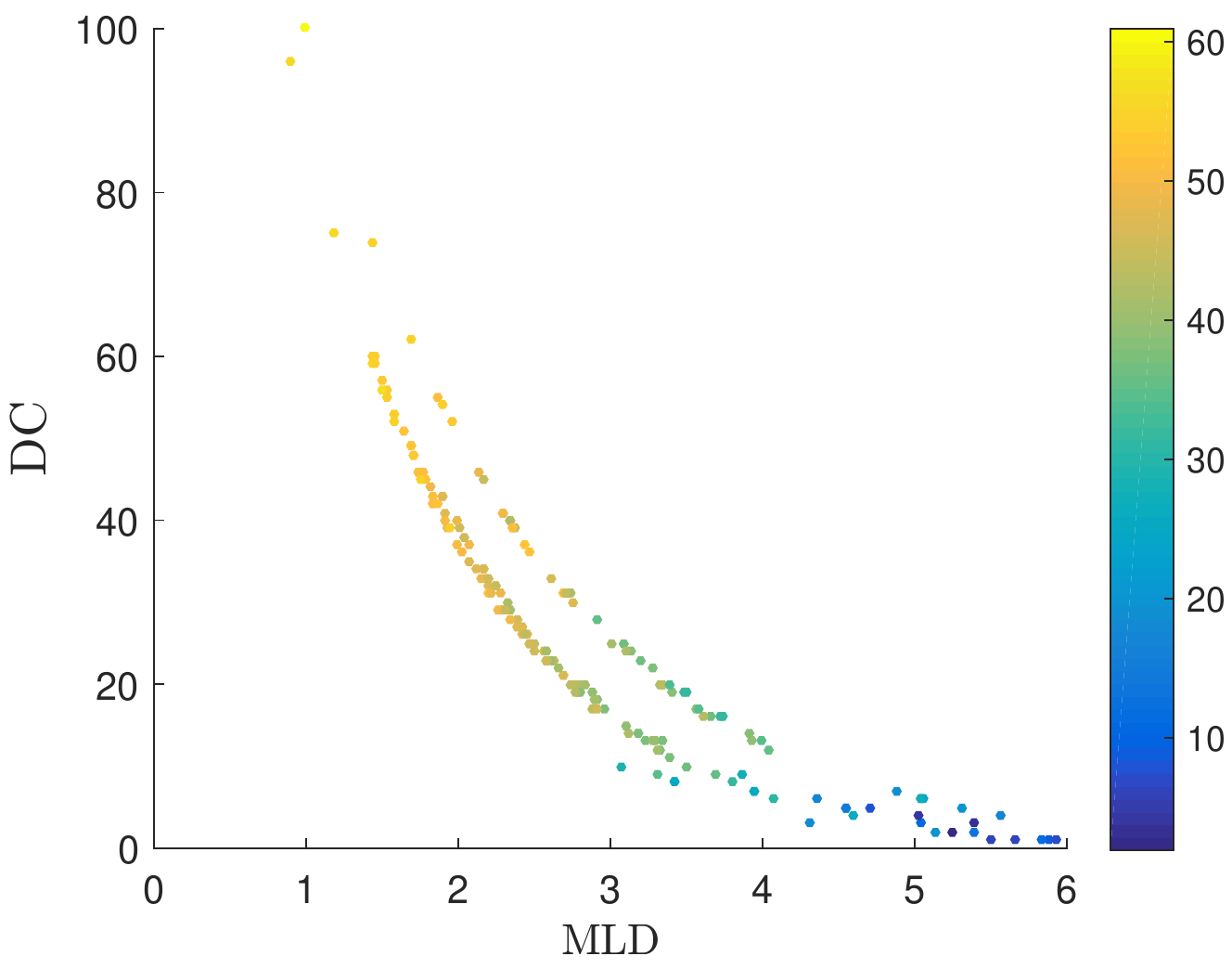}
    \end{minipage}}
  
    \subfigure[USAir network]{
      \label{fig_USAir_DC} 
      \begin{minipage}[b]{0.5\textwidth}
      \centering
      \includegraphics[width=\textwidth]{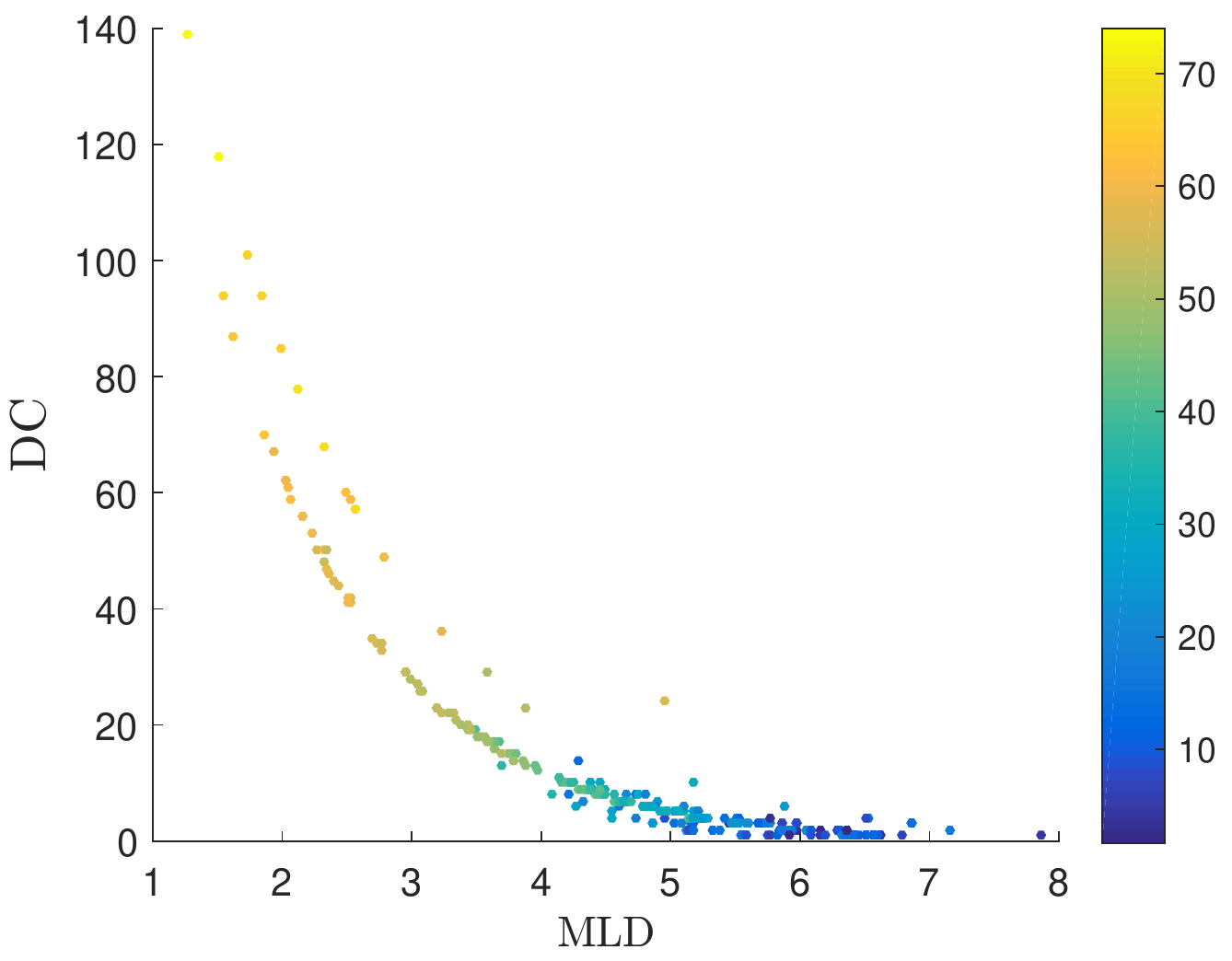}
      \end{minipage}}%
      \subfigure[Political blogs network]{
      \label{fig_Poli_DC} 
      \begin{minipage}[b]{0.5\textwidth}
      \centering
      \includegraphics[width=\textwidth]{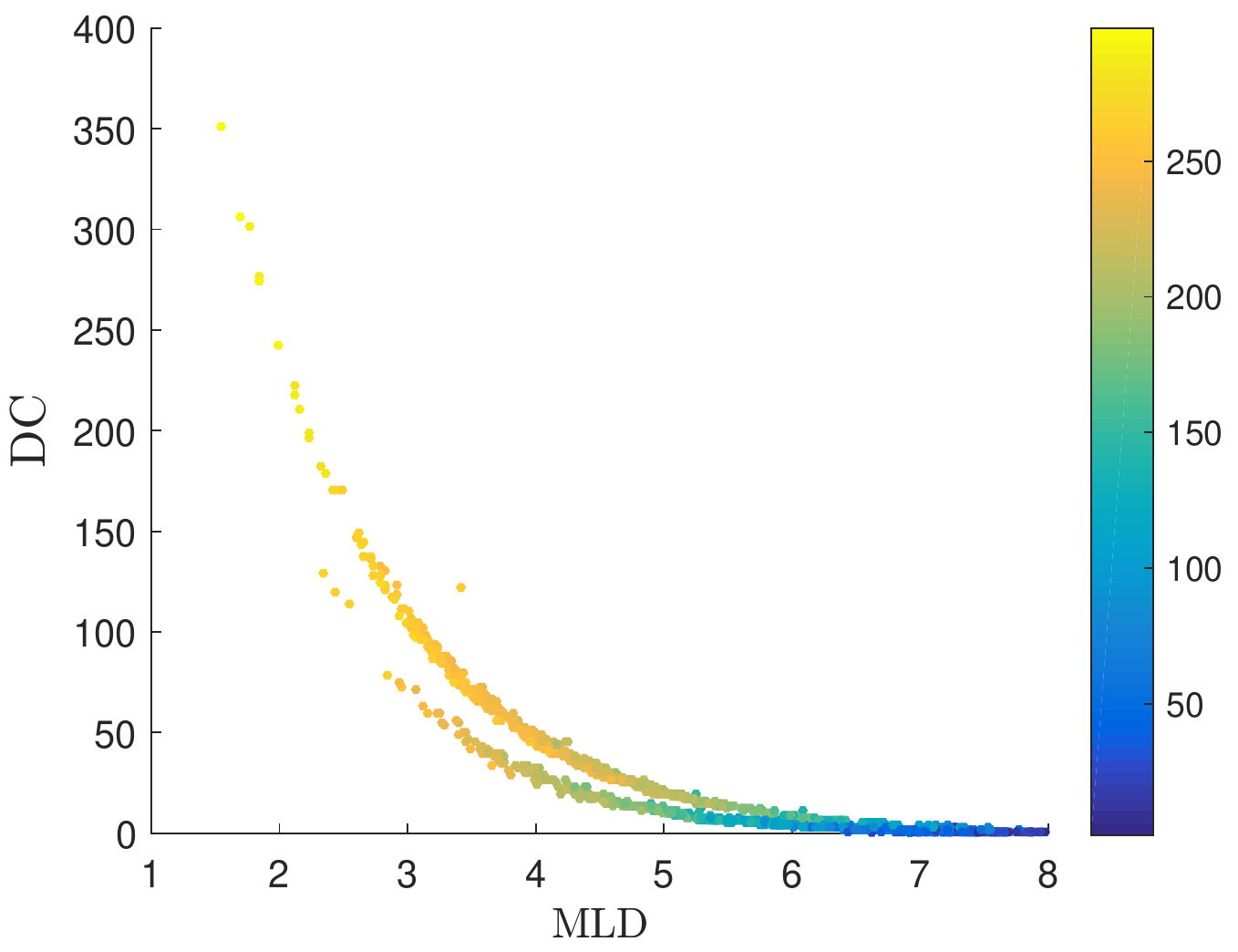}
      \end{minipage}}
  \caption{\textbf{The relationship between MLD and DC when $\lambda = 0.05$ in real-world networks.}The value on the horizontal and vertical axes means the value obtained by MLD and DC respectively, and the color of point means the infectious ability obtained by SI model.}
  \label{fig_COrre_DC} 
  \end{figure}


Because BC has lots of nodes with same value which would cause unusual relationship between BC and this proposed method, the comparison methods are chosen as CC and DC in this section. The relationship between the values obtained by different centrality measures and the infectious ability obtained by SI model are shown in Fig. \ref{fig_COrre_CC} (CC VS MLD) and Fig. \ref{fig_COrre_DC} (DC VS MLD). One point in the relationship graph represents one node in the network, the value of axis means the value obtained by different measures, and the color of point shows the infectious ability of this node obtained by SI model, i.e. the number of infected nodes ($F(10)$) in 10 steps. The infectious ability of node is obtained by averaging 50 independent experiments results when $\lambda = 0.05$. The positive correlation means the nodes would have large value obtained by comparison method and MLD, and negative correlation is the opposite. Observing from Fig. \ref{fig_COrre_CC}, CC and MLD is negative correlation, and their relationship is linear which can give similar rank results between these two measures. In addition, the values obtained by CC is relatively small than other methods (small order of magnitude) which cannot clearly show the difference in nodes' importance. Observing from Fig. \ref{fig_COrre_DC}, the correlation between DC and MLD is also negative, which means the node with large MLD would have small DC. What's more, it can be found that there are lots of nodes with small degree centrality, which is because of the scale-free property of the complex network. Thus, there would be lots of nodes with small DC that cannot correctly identify importance. However, MLD can overcome this shortcoming, because the MLD of node would be more scattered which can give each node with unique value and obtain a relatively reasonable rank lists. Overall speaking, this proposed method would be different withe existing methods, which is negative correlated with exiting methods. In addition, this proposed method can give a more reasonable rank list because it can identify the importance of nodes with close value obtained by existing methods.


\section{Conclusion} \label{Conclusion}

In this paper, a novel method is proposed to identify the influential nodes based on multi-local dimension in the complex networks. Different with previous methods, this proposed method is a more general model, because it can degenerate to local information dimension and variant of local dimension with the different chosen of weighting coefficient $q$. In addition, this proposed method is negative correlated with existing methods which means the influential nodes would have small value of MLD and large value of existing centrality measures. Comparing with exiting centrality methods, this proposed method can effectively identify the influential nodes in the network and give a reasonable rank to these nodes, which can overcome the limitations of previous methods. 

However, this proposed method can still be improved to meet the high requirements in this field. For instance, there are still some nodes with same value of MLD, and the ranking of these nodes is relative top, which can mislead to form the correct node importance rank. Thus, in further research, the consideration factors of this method can be changed, which can demonstrate the property of the network more specifically. 

\section*{Acknowledgment}
The authors greatly appreciate the reviews' suggestions and the editor's encouragement. The work is partially supported by National Natural Science Foundation of China (Grant Nos. 61973332, 61573290, 61503237).

\bibliographystyle{ieeetr}
\bibliography{mybibfile}

\end{document}